\documentclass[aps,prd,showpacs,preprintnumbers,showpacs,twocolumn,groupedaddress]{revtex4}
\usepackage{graphicx}
\usepackage{amsmath}
\usepackage{color}
\newcommand{\Slash}[1]{\ooalign{\hfil/\hfil\crcr$#1$}}
\newcommand{\tr}{\mathrm{tr}}
\newcommand{\Tr}{\mathrm{Tr}}

\begin{document}

\title{Charged vector mesons in a strong magnetic field}
\preprint{RIKEN-QHP-44}
\author{Yoshimasa~Hidaka}
\author{Arata~Yamamoto}
\affiliation{Theoretical Research Division, Nishina Center, RIKEN, Wako 351-0198, Japan}

\date{May 7, 2013}

\begin{abstract}
We show that charged vector mesons cannot be condensed by a magnetic field.
Although some hadron models predict the charged vector meson condensation in a strong magnetic field, we prove, by means of the Vafa-Witten theorem, that this is not the case in QCD.
We also perform the numerical analysis for the meson mass and condensation in lattice QCD.
The lattice QCD data confirm no charged vector meson condensation in a magnetic field.
\end{abstract}


\maketitle

\section{Introduction}
Strong magnetic fields drastically affect dynamics of the strong interaction.
Such strong magnetic fields were/are created in the early Universe, heavy-ion collision experiments, and magnetars~\cite{Vachaspati:1991nm,Enqvist:1993kf,Kharzeev:2007jp,Skokov:2009qp,Deng:2012pc}.
A magnetic field modifies the energy spectrum of particles, in particular, charged particles.
The energy levels of charged particles are quantized, which is the so-called Landau quantization, due to circular motion by the Lorentz force.
In addition to the Landau quantization, the anomalous Zeeman splitting happens for the particles with spin.
The energy levels of free charged particles in a background magnetic field parallel to $z$ axis are given as $E^2 = p_z^2 + (2n+1)|qB| - gs_z qB + m^2$, where $g$ is the $g$-factor.
Masses of charged hadrons are expected to obey this formula in the weak magnetic field limit.
For example, charged pions become heavier in a magnetic field as  $m^2_{\pi^\pm} (B) = m^2_{\pi^\pm} (B=0) + eB$.
On the other hand, the polarized charged $\rho$ mesons become lighter as  $m^2_{\rho^\pm} (B) = m^2_{\rho^\pm} (B=0) - eB$, where the $g$-factor is estimated as $g=2$~\cite{Samsonov:2003hs,Braguta:2004kx,Aliev:2004uj,Bhagwat:2006pu,Hedditch:2007ex,Lee:2008qf}.
Intuitively, the mass of charged $\rho$ mesons becomes zero at the critical magnetic field $eB_c 
\approx m_{\rho^\pm}^2 (B=0)$, and seems imaginary above the critical magnetic field.
The charged vector meson condensation above the critical magnetic field was suggested in hadronic models \cite{Schramm:1991ex,Chernodub:2010qx,Chernodub:2011gs}.
This condensation was also discussed in more microscopic theories, such as the extended Nambu--Jona-Lasinio model (NJL)~\cite{Chernodub:2011mc}, 
the lattice QCD simulation \cite{Braguta:2011hq}, 
and models of the gauge/gravity correspondence \cite{Callebaut:2011uc,Ammon:2011je}.

In this paper, we point out that the charged vector meson condensation cannot occur in QCD in a strong magnetic field.
More generally, any global-internal symmetry is not spontaneously broken by a magnetic field. 
For the vanishing magnetic field, this is known as the Vafa-Witten theorem~\cite{Vafa:1983tf},
which is a consequence of the positivity of the fermion determinant. 
The positivity is maintained even in the existence of a magnetic field, so that the Vafa-Witten theorem works.

We also study how the charged vector meson mass depends on a magnetic field in lattice QCD.
Meson masses in a magnetic field have been calculated in a few cases \cite{Lee:2008qf,Bali:2011qj,Luschevskaya:2012xd}.
When the strength of a magnetic field exceeds the QCD scale, internal structures of hadrons are important.
Lattice QCD is the best way to study hadron properties in a strong magnetic field quantitatively.

This paper is organized as follows: In Sec.~\ref{sec:Theoretical analysis}, we  discuss the possibility of the charged vector meson condensation,
and analytically show that the charged vector meson condensation does not occur in a magnetic field.
In Sec.~\ref{sec:NumericalAnalysis}, we evaluate the meson masses in a magnetic field using lattice QCD and numerically confirm the condensation does not occur.
Section~\ref{sec:Summary} is devoted to a summary and discussion.

\section{Theoretical analysis}\label{sec:Theoretical analysis}
\subsection{Symmetry of QCD in a magnetic field}
We consider two-flavor QCD in a magnetic field.
The up- and down- quark masses are the same and nonzero, $m \equiv m_u = m_d \ne 0$.
We assume that the strong $CP$ angle is zero.
We work in Euclidean space, and choose $\gamma_\mu^\dag=\gamma_\mu$ and $\{\gamma_\mu,\gamma_\nu\}=2\delta_{\mu\nu}$.
The Lagrangian reads
\begin{equation}
\mathcal{L}=\frac{1}{2}\tr G_{\mu\nu}G_{\mu\nu} +\bar{\psi}(\Slash{D} +m)\psi,
\end{equation}
with the QCD field strength $G_{\mu\nu}=-i[D_\mu,D_\nu]/g_{\rm YM}$.
The Dirac operator is 
\begin{equation}
D_\mu=\partial_\mu -ig_{\rm YM}A_\mu-iq A^{\text{em}}_\mu
\end{equation}
with the electric charge $q=e(\tau_3+1/6)$.

When $A^{\text{em}}_\mu=0$ for all $\mu$, this Lagrangian has an internal global $SU(2)_V\times U(1)_B$ symmetry,
in addition to space-time (Poincar\'e)  and discrete ($C$, $P$, and $T$) symmetries.
Chiral symmetry is explicitly broken by the quark mass.

When a constant external magnetic field exists along the $z$-direction, parts of these symmetries are explicitly broken.
The global symmetry is broken down into $U(1)_B\times U(1)_{I_3}$. $C$ and $T$ are broken while $P$ is preserved.
Lorentz symmetry is also broken down into $SO(1,1)_{t,z}\times SO(2)_{x,y}$.

\subsection{Vafa-Witten theorem}

Here, we discuss the possibility of the charged vector meson condensation in a magnetic field following
Weinberg's textbook~\cite{Weinbergtext}.
The order parameters of the charged vector meson condensation are $\langle\bar{\psi}\tau_+\gamma_+\psi\rangle$ and $\langle\bar{\psi}\tau_-\gamma_-\psi\rangle$ with $\gamma_\pm \equiv (\gamma_1 \pm i \gamma_2)/2$ and $\tau_\pm \equiv (\tau_1 \pm i \tau_2)/2$.
We will show that this type of the operator cannot condense in QCD (more generally, vectorlike gauge theories).
In the vanishing magnetic field, this theorem is called the Vafa-Witten theorem~\cite{Vafa:1983tf}.

Since the Dirac operator in Euclidean space is anti-Hermitian,
the eigenvalues of the Dirac operator are pure imaginary.
In addition, the Dirac operator anticommutes with $\gamma_5$, which implies that 
the Dirac operator has a pair of eigenvalues $\pm i\lambda$ for nonzero $\lambda$. This can be shown in the following.
Suppose that $\psi_\lambda$ is the eigenstate of the Dirac operator with an eigenvalue $i\lambda$ ($\lambda$ is real), which satisfies the eigenvalue equation, $\Slash{D} \psi_\lambda =  i\lambda \psi_\lambda$.
Multiplying both sides of the eigenvalue equation by $-\gamma_5$, one obtains $\Slash{D} \gamma_5\psi_\lambda =  -i\lambda \gamma_5\psi_\lambda$.
Therefore, $\gamma_5\psi_\lambda$ is the eigenstate with an eigenvalue $-i\lambda$. 
This fact implies that the fermion determinant is real and positive:
\begin{equation}
\begin{split}
\det (\Slash{D}+m)&=\prod_{\lambda}(i\lambda+m)\\
&=m^{n_0}\prod_{\lambda>0}(i\lambda+m)(-i\lambda+m)\\
&=m^{n_0}\prod_{\lambda>0} (\lambda^2+m^2)>0 \ ,
\end{split}
\end{equation}
where $n_0$ is the number of zero eigenvalues of the Dirac operator.
By integrating the fermion degrees of freedom, the effective measure is obtained as
\begin{equation}
d\mu= \prod_{\mu,a,x}dA_\mu^a(x)\det(\Slash{D}+m) e^{-S[A]} \ ,
\end{equation}
which is also real and positive.
Here, $S[A]$ is the action of the gauge field.
We define the gauge average with the effective measure as
\begin{equation}
\langle \mathcal{O}\rangle_A\equiv \frac{1}{\int d\mu}\int d\mu\ \mathcal{O} \ .
\end{equation}
Since the effective measure is real and positive, $|\langle \mathcal{O}\rangle_A|\leq \langle |\mathcal{O}|\rangle_A$,
which plays important role in the QCD inequalities.

Another important property of the Dirac operator is that the following operator is bounded by the bare quark mass:
\begin{equation}
\begin{split}
\frac{1}{\Slash{D}+m}\frac{1}{\Slash{D}^\dag+m}&=\frac{1}{-\Slash{D}^2+m^2}\\
&=\sum_\lambda \frac{1}{\lambda^2+m^2}|\lambda\rangle\langle \lambda|\\
&\leq \sum_\lambda \frac{1}{m^2}|\lambda\rangle\langle \lambda|=\frac{1}{m^2} \ ,
\label{eq:quarkBound}
\end{split}
\end{equation}
which is independent of the gauge field. In other words, the operator norm of $(\Slash{D}+m)^{-1}$ is equal to $1/m$.

Now, let us consider the symmetry breaking.
We write the order parameter as
\begin{equation}
\phi\equiv \frac{1}{\mathcal{N}}\int d^4x \, \bar{\psi}(x)F\psi(x) \ ,
\end{equation}
where $\mathcal{N}=4VV_pN_cN_f$ is a normalization factor with $N_c=3$, $N_f=2$, and $V$ ($V_p$) being the space-time (momentum space) volume. 
We keep $V_p$ finite to make the calculation well defined: $V_p\sim 1/a^4$, in the Lattice regularization with a lattice spacing $a$.
We assume that symmetry breaking is independent of the choice of regularization schemes  since condensation is triggered by infrared dynamics.
The operator $F$ depends on  the isospin and spinor, which is normalized as $\Tr F F^\dag =1$.
Here, $\Tr$ denotes the sum over space-time, isospin, and spinor indices and is normalized to unity.
In order to take into account the possibility of an inhomogeneous phase, $F$ may depend on the space-time coordinate.
For example, if one considers an charge density wave with a single-plane wave,  $\langle \psi^\dag{\psi}\rangle=\Delta \cos\theta(x)$,
one may choose $F=\sqrt{2}\gamma^0\cos\theta(x)$; then $\langle\phi\rangle=\Delta/(4N_cN_f\sqrt{2})$.
For the charged vector meson condensation, we can choose $F= \tau_+\gamma_+f(x)$, where $f(x)$ is the function characterizing inhomogeneity.

In the following, we consider the order parameter satisfying
\begin{equation}
\Tr \,F\frac{1}{\Slash{D}+m}=0 \ .
\label{eq:vanishingorderparameter}
\end{equation}
This is  the case with the condensation of the charged-vector meson
since it carries the electric charge and its trace vanishes.
It cannot, however, be concluded that the symmetry breaking does not occur because this vacuum might be unstable for
a small disturbance.
To make the discussion of the symmetry breaking precise, we need to add an explicit breaking term $\epsilon \bar{\psi}\varGamma\psi$ into the Lagrangian,
and we take $\epsilon\to0$ after all the averages. 
Here, $\varGamma$ depends on isospins, spinors, and the space-time coordinate.
We assume that $\varGamma$ is anti-Hermitian,
which ensures that  the effective measure is positive definite.
We also assume that the operator norm of $\varGamma$, $C\equiv \|\varGamma\|_\text{op}$, is bounded ($C<\infty$).
Then, the expectation value of $\phi$ with the explicit breaking term becomes
\begin{equation}
\begin{split}
\langle \phi \rangle_\epsilon = \langle \Tr F\frac{1}{\Slash{D}+m+\epsilon \varGamma} \rangle_{A,\epsilon} \,.
\label{eq:TrF}
\end{split}
\end{equation}
If $\lim_{\epsilon\to0} \lim_{V\to\infty}\langle \phi \rangle_\epsilon$ is nonzero, the symmetry is spontaneously broken.
Here, we show $\lim_{\epsilon\to0} \lim_{V\to\infty}\langle \phi \rangle_\epsilon=0$.
To see this, let us expand $\Tr F(\Slash{D}+m+\epsilon \varGamma)^{-1}$ in terms of $\epsilon$,
\begin{equation}
\begin{split}
 \Tr F\frac{1}{\Slash{D}+m+\epsilon \varGamma}=\sum_{n=1}^\infty (-1)^n\epsilon^n \Tr F\frac{1}{\Slash{D}+m} \left(\varGamma\frac{1}{\Slash{D}+m} \right)^n,
\end{split}
\end{equation}
where we dropped the $n=0$ term using Eq.~(\ref{eq:vanishingorderparameter}).
Using the H\"{o}lder inequality,
$|\Tr \,\mathcal{O}_1 \mathcal{O}_2|\leq\|\mathcal{O}_1\|_\text{op} \Tr\, \sqrt{\mathcal{O}_2 \mathcal{O}_2^\dag}$ for linear operators, $\mathcal{O}_1$ and  $\mathcal{O}_2$,
 we obtain the absolute value of the coefficient as 
\begin{equation}
\begin{split}
&\left|(-1)^n \Tr\, F\frac{1}{\Slash{D}+m} \left(\varGamma\frac{1}{\Slash{D}+m} \right)^n\right|\\
&\quad\leq
\left\|\frac{1}{\Slash{D}+m} \left(\varGamma\frac{1}{\Slash{D}+m} \right)^n\right\|_\text{op}
\Tr\, \sqrt{FF^\dag}\\
&\quad\leq
\| \varGamma\|_\text{op}^n\left\|\frac{1}{\Slash{D}+m}\right\|_\text{op}^{n+1}
\sqrt{\Tr\, FF^\dag}
=\frac{C^{n}}{m^{n+1}}\,,
\end{split}
\end{equation}
where $\| (\Slash{D}+m)^{-1}\|_\text{op}=1/m$ is used.
The series is absolute convergent if $m> \epsilon C $.
Therefore, Eq.~(\ref{eq:TrF}) is bounded for a small perturbation
as
\begin{equation}
\begin{split}
\left| \Tr\, F\frac{1}{\Slash{D}+m+\epsilon \varGamma}\right| \leq \sum_{n=1}^\infty \frac{(\epsilon C)^n}{m^{n+1}}=\frac{\epsilon C}{m}\frac{1}{m-\epsilon C} \,.
\end{split}
\end{equation}
This bound is independent of gauge configurations and  the space-time volume, so that it does not change by the average of the gauge field and the large volume limit.
Since the bound of $\langle\phi\rangle_{\epsilon}$ is an analytic function of $\epsilon$, we can smoothly take $\epsilon\to0$.
In this limit, $\langle\phi\rangle_\epsilon$ vanishes;
therefore, the charged vector meson condensation does not occur in QCD.

More generally, according to Vafa-Witten's argument~\cite{Vafa:1983tf}, one can show that  no Nambu-Goldstone boson exists in a magnetic field if $m\neq0$.
In other words, any internal symmetry cannot be spontaneously broken down in a magnetic field with a nonzero fermion mass as long as (discrete) translational symmetry is not broken at least in one direction.

We note that this argument cannot apply to the vectorlike theory in which the vector field, $V^a_{\mu}$, has an electric charge, like the extended NJL model~\cite{Ebert:1982pk,Ebert:1985kz}.
In this case, the given background carries the electric charge, so that 
\begin{equation}
\Tr \frac{1}{\Slash{D}+m}F
\end{equation}
may not vanish, where $D_{\mu}=\partial_\mu-i\tau^aV^a_{\mu}-iqA_\mu^{\rm em}$.
Therefore, the previous argument cannot apply, and the charged vector meson condensation cannot be excluded.
In fact, the charged vector meson condensation is found in the extended NJL model~\cite{Chernodub:2011mc}.

\subsection{QCD inequality}
In the previous subsection, we discussed that the charged vector meson condensation does not occur.
Here, we consider the lower bound of the charged vector meson mass.
For this purpose, we apply the so-called QCD inequalities~\cite{Weingarten:1983uj,Witten:1983ut,Nussinov:1983hb,Espriu:1984mq,Nussinov:1999sx}, which follows that
\begin{equation}
\begin{split}
& \langle \rho^-(x)\rho^+(y) \rangle \\
&=-\langle \tr S_u(x,y)\gamma_+ S_d(y,x) \gamma_- \rangle_A\\
&\leq \sqrt{\langle\tr S_u(x,y) S^\dag_u(x,y)\rangle_A \langle  \tr S_d(x,y) S^\dag_d(x,y)\rangle_A} \ ,
\end{split}
\end{equation}
where we have used the Cauchy-Schwartz inequality in the second line.
Here, $S_u(x,y)$ and $S_d(x,y)$ are the propagators of up and down quarks, respectively.
Thanks to $\gamma_5$ Hermiticity of the Dirac operator, $\Slash{D}^\dag = \gamma_5\Slash{D}\gamma_5$,
it follows that
\begin{equation}
\begin{split}
&\langle\tr S_u(x,y) S^\dag_u(x,y)\rangle_A \\
&=\langle\tr S_u(x,y) \gamma_5S_u(y,x)\gamma_5\rangle_A\\
&=\langle \bar{u}(x)i\gamma_5 u(x)
\bar{u}(y)i\gamma_5 u(y)
\rangle_\text{conn.} \ .
\end{split}
\end{equation}
We will call this correlation function in the last line
the ``connected'' neutral pion, $\pi^\text{c}_u$ (and $\pi^\text{c}_d$ for the down quark).
At large distance, $\langle \rho^-(x) \rho^+(y)\rangle\sim \exp(-|x-y|m_{\rho^+})\leq\exp(-|x-y|(m_{\pi^\text{c}_u}+m_{\pi^\text{c}_d} )/2)$.
Therefore, $m_{\rho^\pm}\geq (m_{\pi^\text{c}_u}+m_{\pi^\text{c}_d} )/2\geq\mathrm{min}(m_{\pi^\text{c}_u},m_{\pi^\text{c}_d} )$.
Although the connected neutral pions are not physical neutral pions, they can be calculated in the lattice QCD, as discussed in the next section.

\section{Numerical analysis}\label{sec:NumericalAnalysis}
\subsection{Simulation setups}
We performed the quenched QCD simulation with $\beta = 5.9$.
For the quark propagator, we used the Wilson fermion with the hopping parameter $\kappa = 0.1583$.
These parameters correspond to the lattice spacing $a \simeq 0.10$ fm and the meson mass ratio $m_\pi / m_\rho \simeq 0.59$ \cite{Aoki:2002fd}.
A constant Abelian magnetic field is applied in the $z$-direction \cite{AlHashimi:2008hr}.
The masses of up and down quarks are the same, but the electric charges are different as $q = {\rm diag} ( q_u, q_d ) = {\rm diag} ( 2e/3, -e/3 )$.
To suppress unnecessary contributions from finite-momentum excited states, all the correlation functions were projected onto $p_z = 0$ by averaging the positions of the source operators in the $z$-direction.
Unlike the usual meson mass calculation without the magnetic field, the correlation functions were not projected onto $p_x = p_y = 0$ because the background Abelian gauge field breaks translational invariance in the $x$- and $y$-directions.

We calculated the correlation functions
\begin{eqnarray}
G_X(t_1-t_2) = \langle X^\dagger (\vec{x},t_2) X (\vec{x},t_1) \rangle
\end{eqnarray}
of four mesons $X = \{\pi^+, \rho^+, \pi^0,\rho^0 \}$.
For charged and neutral $\rho$ mesons, we used the correlation function
\begin{eqnarray}
G_{\rho}(t_1-t_2) = \frac{1}{2} \sum_{\mu=1,2} \langle \rho^\dagger_\mu (\vec{x},t_2) \rho_\mu (\vec{x},t_1) \rangle
\end{eqnarray}
with $\rho_\mu^+ = \bar{\psi} \gamma_\mu \tau_+ \psi$ and $\rho_\mu^0 = \bar{\psi} \gamma_\mu \tau_3 \psi$ .
This correlation function couples to both of the polarized and antipolarized (i.e., $s_z=\pm 1$) components of $\rho$ mesons.
Only the lowest energy state survives in the large $t$ limit.
Even if we do not know which component is the lowest energy state, especially of the $\rho^0$ meson, we can automatically extract the mass of the lowest energy state from this correlation function.

We did not calculate the $\mu =3$ (i.e., $s_z=0$) components of $\rho$ mesons. 
It is too difficult to calculate them in lattice QCD.
In the background magnetic field, the $\pi$-$\rho_{3}$ mixing exists even for in the connected diagram.
Thus, the $\mu =3$ component of a $\rho$ meson is an excited state of a pion.
At least in the weak magnetic field limit, there is a large number of magnetic-splitting states of the pion below the energy level of the $\rho$-meson state.
We cannot calculate such a highly excited state in the lattice QCD simulation.

For neutral $\pi$ and $\rho$ mesons, we calculated only the connected diagram, which is necessary for the QCD inequality.
While the disconnected diagram is forbidden in the absence of the magnetic field, it is allowed in the presence of the magnetic field because the magnetic field breaks isospin symmetry.
We ignored the disconnected diagram in this simulation.
In this sense, our neutral mesons are not physical ones.

\subsection{Meson masses}
\begin{figure}[h]
\begin{center}
\includegraphics[scale=1]{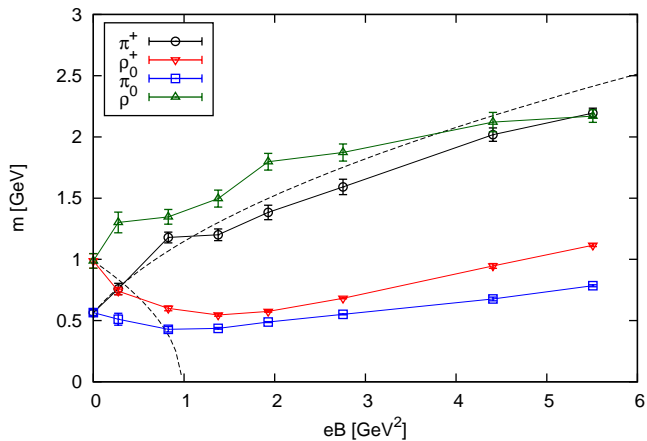}
\caption{\label{L1}
The meson masses in a magnetic field.
The broken curves are $m^2_{\pi^+} (B) = m^2_{\pi^+} (B=0) + eB$ and $m^2_{\rho^+} (B) = m^2_{\rho^+} (B=0) - eB$.
}
\end{center}
\end{figure}

We performed the standard mass analysis of ground-state mesons in lattice QCD.
The meson masses were extracted from the fitting function
\begin{equation}
G_X(t) = A_X \cosh [m_X(t-aN_t/2)]
\end{equation}
in large $t$.
The lattice volume is $N_s^3 \times N_t = 16^3 \times 32$.
The numerical results are shown in Fig.~\ref{L1}.

The charged pion mass increases in the magnetic field.
This mass shift can be explained by the naive mass formula $m^2_{\pi^+} (B) = m^2_{\pi^+} (B=0) + eB$.
As shown in the figure, this formula well reproduces the present lattice result in a weak magnetic field.
This behavior was also observed in the full QCD simulation \cite{Bali:2011qj}.
The lattice data slightly deviate from this formula in a strong magnetic field.

The charged $\rho$ meson mass shows a nontrivial dependence on the magnetic field.
When the magnetic field is weak, the mass is a decreasing function of the magnetic field.
The naive mass formula, $m^2_{\rho^+} (B) = m^2_{\rho^+} (B=0) - eB$, reproduces the lattice data.
At $eB\simeq 1$ GeV$^2$, the mass has a nonzero minimum.
When the magnetic field is stronger than this value, the mass becomes an increasing function of the magnetic field.
As a consequence, the charged $\rho$ meson is always massive and heavier than the connected neutral pion in the whole range of the magnetic field.
Although the Wilson fermion does not have the exact positivity, the present lattice result is consistent with the Vafa-Witten theorem and the QCD inequality.

The neutral mesons are much more nontrivial.
In the naive mass formula, neutral particles are independent of a magnetic field.
The lattice result suggests, however, that the neutral meson masses depend on the magnetic field.
This is due to the internal structure of the mesons.
To know how the physical neutral mesons behave in a magnetic field, we have to take into account the disconnected diagram.

When the magnetic field is extremely strong, i.e., $eB \gg 1$ GeV$^2$, the masses of all the mesons monotonically increase.
This is interpreted as a sign that the internal quarks obtain the large magnetic-induced mass.
The underlying mechanism is unknown in the present analysis.

\subsection{Meson condensations}
To exclude the possibility of the charged $\rho$ meson condensation in lattice QCD, we performed another analysis.
If a meson condensation exists, the ground state becomes massless and a long-range correlation appears.
The correlation function becomes
\begin{equation}
G_X'(t) = A_X \cosh [m_X(t-aN_t/2)] + C_X
\label{eqGfit2}
\end{equation}
in large $t$.
If the constant parameter $C_X$ is finite in the limit $N_t\to \infty$, $C_X$ corresponds to the squared meson condensation $\langle X \rangle^2$, and $m_X$ corresponds to the mass of the first excited state.
A similar analysis was performed in a previous work \cite{Braguta:2011hq}.
However, such a constant term can be easily generated by a finite-volume artifact.
We must carefully check the finite-volume artifact.
In particular, we need a larger size in the fitting direction, i.e., in the $t$-direction in Eq.~(\ref{eqGfit2}), because $C_X$ coincides $\langle X \rangle^2$ only in the limit $N_t\to \infty$.

\begin{figure}[h]
\begin{center}
\includegraphics[scale=1]{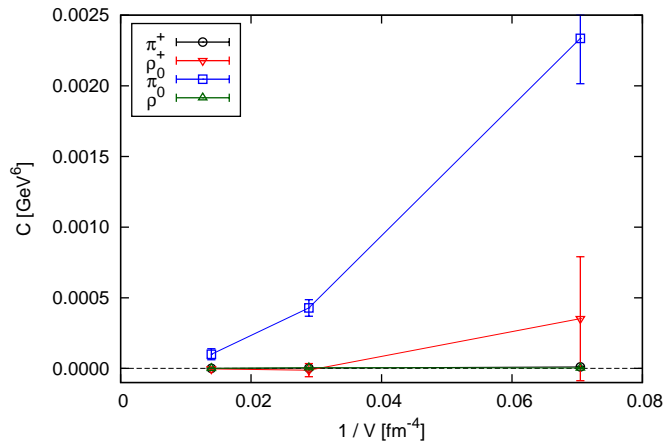}
\caption{\label{L2}
The volume dependence of the constant parameter $C_X$ in Eq.~(\ref{eqGfit2}).
The magnetic field is $eB \simeq 4.3$ GeV$^2$.
}
\end{center}
\end{figure}

We calculated the correlation functions $G_X(t)$ with three lattice volumes $N_s^3 \times N_t = 16^3 \times 32$, $20^3 \times 40$ and $24^3\times 48$, and fitted the results with Eq.~(\ref{eqGfit2}).
The numerical settings are summarized in Table \ref{tab1}.
In Fig.~\ref{L2}, we show $C_X$ as a function of the lattice volume $V = a^4 N_s^3 N_t$.
The magnetic field is fixed at a large value $eB \simeq 4.3$ GeV$^2$.
In a small volume, $C_{\pi^0}$ and $C_{\rho^+}$ seem finite.
In the infinite volume limit, however, all $C_X$ approach to zero.
In particular, $C_{\rho^+}$ is zero within the statistical error.
From this analysis, we conclude that the charged $\rho$ meson is not condensed by a magnetic field.

As shown in Fig.~\ref{L2}, $C_{\pi^0}$ is large compared to other mesons.
This is an expected behavior because the connected neutral pion is the lightest particle and the finite-volume artifact is the largest for the lightest particle.
If there were the physical charged $\rho$ meson condensation, $C_{\rho^+}$ would be finite and larger than $C_{\pi^0}$.
To estimate the finite-volume artifact of a small quantity, it is important to compare it with the other quantity that is the most sensitive to the finite-volume artifact.

\begin{table}[h]
\renewcommand{\tabcolsep}{0.25pc}
\caption{\label{tab1}
The numerical settings in Fig.~\ref{L2}.
In fitting with Eq.~(\ref{eqGfit2}), the fit range $t_{\rm fit}$, the best-fit value of $C_{\rho^+}$, and $\chi^2/$d.o.f.(number of degrees of freedom) are listed.
}
\begin{center}
\begin{tabular}{ccccccc}
\hline\hline
$eB$ [GeV$^2$] & $a$ [fm] & $N_s$ & $N_t$ & $t_{\rm fit}$ [a] & $C_{\rho^+}$ [GeV$^6$] & $\chi^2/$d.o.f. \\
\hline
4.3 & 0.10 & 16 & 32 & 8-24 & 35(44)$\times 10^{-5}$ & 1.43 \\
4.3 & 0.10 & 20 & 40 &10-30 & 13(46)$\times 10^{-6}$ & 0.73 \\
4.3 & 0.10 & 24 & 48 &12-36 & 33(77)$\times 10^{-7}$ & 1.45 \\
\hline\hline
\end{tabular}
\end{center}
\end{table}

\section{Summary and Discussion}\label{sec:Summary}
We have analytically and numerically shown that charged vector mesons are not condensed by a magnetic field.
In general, a magnetic field cannot induce any condensation that breaks global-internal symmetry in QCD.
In other words, a strong magnetic field cannot change the QCD vacuum structure.
For example, the insulator-superconductor phase transition cannot be induced by a strong magnetic field alone.

The Vafa-Witten theorem works not only in the magnetic field but also at finite temperature, $T$, 
so that we can conclude that no Nambu-Goldstone phase associated with spontaneous breaking of global-internal symmetries
exists in the phase diagram of the $(T,B)$ plane. 
We note that we cannot exclude the possibility of spontaneous breaking of space-time symmetries because the order parameter
does not satisfy Eq.~(\ref{eq:vanishingorderparameter}), i.e., the disconnected diagram exists.

As expected in hadron models, if a charged vector meson were a point particle, its mass would decrease in a magnetic field.
The lattice QCD results support this scenario in the weak magnetic field limit.
The mass of the charged vector meson turns to increase at $eB \sim 1$ GeV$^2$, which is the QCD scale, before the mass reaches zero.
When the magnetic field is stronger than this scale, other meson masses also increase monotonically,
where the internal structure of the mesons becomes non-negligible and the validity of hadron models break down.
Therefore, we expect that some transition occurs from a hadronic phase
to a phase governed by the magnetic scale although this is not a phase transition separated by some symmetries.
A possible scenario to explain the increasing masses is 
that the constituent quark mass increases by the magnetic catalysis~\cite{Suganuma:1990nn,Klimenko,Gusynin:1994re,Shushpanov:1997sf,Shovkovy:2012zn}.
In this case, the meson masses increase as $\sqrt{eB}$.
The detailed analysis for this transition is beyond our scope in this paper.

\section*{ACKNOWLEDGMENTS}
The authors thank Masanori Hanada, Yuji Sakai, and Naoki Yamamoto for useful discussions.
A.~Y.~is supported by the Special Postdoctoral Research Program of RIKEN.
This work was supported by JSPS KAKENHI Grant Numbers 23340067 and 24740184.
The lattice QCD simulations were carried out on NEC SX-8R in Osaka University.

\end{document}